\documentclass{ws-ijmpd}
\usepackage{graphicx}  

\newcommand{\fin}{f_{in}}
\newcommand{\fo}{f_{out}}
\newcommand{\fot}{f_{12}}
\newcommand{\fto}{f_{21}}
\newcommand{\dt}{\frac{d\,t}{d\tau}}
\newcommand{\dto}{\frac{d\,t}{d\tau_1}}
\newcommand{\dtt}{\frac{d\,t}{d\tau_2}}
\newcommand{\dr}{\frac{d\,r_0}{cd\tau}}
\newcommand{\dro}{\frac{d\,r_1}{cd\tau_1}}
\newcommand{\drt}{\frac{d\,r_2}{cd\tau_2}}
\newcommand{\drot}{\frac{d\,r_1}{cdt}}
\newcommand{\drtt}{\frac{d\,r_2}{cdt}}
\newcommand{\ep}{\epsilon}
\newcommand{\mo}{m_{out}}
\newcommand{\mi}{m_{in}}
\newcommand{\mot}{m_{12}}
\newcommand{\mto}{m_{21}}
\newcommand{\ei}{e_{in}}
\newcommand{\ga}{Gm_1m_2-e_1e_2}
\begin{document}

\markboth{Pizzi and Paolino}
{Intersections of self-gravitating charged shells}

%
\catchline{}{}{}{}{}
%

\title{Intersections of self-gravitating charged shells in a Reissner-Nordstrom field}
\author{M. PIZZI}

\address{Physics Department, Rome University ``La Sapienza'',\\
Piazzale A. Moro,  00185 Rome, Italy, and\\ ICRANet, Pescara, Italy 65122.
\\
pizzi@icra.it}

\author{A. PAOLINO}

\address{Physics Department, Rome University ``La Sapienza'',\\
Piazzale A. Moro,  00185 Rome, Italy.}

\maketitle

\begin{history}
\received{Day Month Year}
\revised{Day Month Year}
\comby{Managing Editor}
\end{history}
\maketitle

\begin{abstract}

 We describe the equation of motion of two charged spherical shells with tangential pressure in the field of a central Reissner-Nordstrom (RN) source. We solve the problem of determining the motion of the two shells \textsl{after} the intersection by solving the related Einstein-Maxwell equations and by requiring a physical continuity condition on the shells velocities. 
 
We consider also four applications: post-Newtonian and ultra-relativistic approximations, a test-shell case, and the ejection mechanism of one shell.
 
 This work is a direct generalization of Barkov-Belinski-Bisnovati-Kogan paper.

\end{abstract}
\section{Introduction}

The mathematical model that we analyze in this paper describes the dynamic evolution of two spherical shells of charged matter which freely move outside the field of a central Reissner-Nordstrom (RN) source. Microscopically these shells are assumed to be composed by charged particles which move on elliptical orbits with a collective variable radius. The angular motion, distributed uniformly and isotropically on the shell surfaces, is mathematically described by a tangential-pressure term in the energy momentum tensor of the Einstein equations. The definition of the shell implies that all the particles have the same following three ratios: energy/mass, angular momentum/mass, and charge/mass. Indeed, since the equations of motion for any singled-out particle ``a'' are
\begin{align}
&	\frac{dt_a}{ds}= \frac{1}{-m_a c^2 g_{tt}(r_a)}(E_a+e_a A_0(r_a))            \label{mot1} \\
&  \left(\frac{dr_a}{ds}\right)^{2}=\frac{1}{m_a^2 c^4}(E_a+e_a A_0(r_a))^2\left(\frac{1}{-g_{tt}(r_a)g_{rr}(r_a)}\right)  - \left(\frac{l_a^2}{m_a^2c^2}\frac{1}{r^2}+1\right) \frac{1}{g_{rr}(r_a)}               \label{mot2}\\
&  \left(\frac{d\theta_a}{ds}\right)^2=\frac{l_a^2}{m_a^2c^2}\frac{1}{r^4}-\frac{k_a^2}{m_a^2c^2}\frac{1}{r^4 \sin^2\theta_a} \label{mot3}\\
&  \frac{d\varphi_a}{ds}=\frac{k_a}{m_a c}\frac{1}{r^2 \sin^2\theta_a}
\end{align}
($g_{tt}$ and $g_{rr}$ are the components of a spherical symmetric metric and $A_0$ is the electric potential; $k_a$ and $l_a$ are arbitrary constants), it is easy to see that the radial motion for all particles is the same if
\begin{align}
	\frac{E_a}{m_a}=const, \ \ \ \ \ \frac{e_a}{m_a}=const, \ \ \ \ \  \frac{|l_a|}{m_a}=const, \ \ \ \ \  \forall a, 
\end{align}
where each $const.$ does not depend on the index $a$. Therefore, if at the beginning the particles are on the same radius $r_a=R_0$, then the shell will evolve ``coherently'', i.e. all particles will evolve with the same radius.

 Now the problem we are interested in is to find the exchange of energy between the two shells after the intersection. Indeed the motion of the shells before and after the crossing can be easily deduced from the equation of motion for just one shell, which equation has been found many years ago by Chase\cite{Cha} with a geometrical method first used by Israel\cite{Isr}. Instead, the intersection problem was considered first by Langer-Eid\cite{LE}: they applied it to the particular case of
electrically neutral and pressureless shells (neutral dust), then Barkov et al.\cite{BBB} considered the more general case of shells intersections when the shells have also tangential pressure (so Langer-Eid's results follows from Barkov et al. results as particular case). Now we generalize this problem for the shells with tangential pressure and also electric charge.
  
  What we achieve in the present paper is the determination of the constant parameters after the intersection knowing just the parameters before the intersection. Actually the unknown parameter is only one, $\mto$, which is the Schwarzschild mass parameter measured by an observer between the shells after the intersection. This parameter is strictly related to the energy transfer which takes place in the crossing, and it is found imposing a proper continuity condition on the shells velocities.

 In the model we assume that there are no other interactions between the two shells apart the gravitational and electrostatic ones. In particular the shells, during the intersection, are assumed to be ``transparent'' each other (i.e. no scattering processes).

The paper is divided as follows: in Sec.2 we preliminarily discuss the one-shell case; in Sec.3, which is the central part of this article, we find the unknown parameter $m_{21}$; then, Secs.4-7 are devoted to some applications: post-Newtonian approximation, zero effective masses case (i.e. ultra-relativistic case), test-shell case, and finally the ejection mechanism.
 
In this paper we deal only with the mathematical aspects of the problem; some astrophysical applications of charged shells in the field of a RN black hole have been considered in Ref.[\refcite{CRV}].

\section{A gravitating charged shell with tangential pressure}
The motion of a thin charged dust-shell with a central RN singularity was firstly studied by De La Cruz and Israel\cite{DI}, while the case with tangential pressure was achieved by Chase\cite{Cha} in 1970. All these authors used the extrinsic curvature tensor and the Gauss-Codazzi equations. However we followed a different way, indeed the same solution can be found also by using $\delta$ and $\theta$ distributions and then by direct integration of the Einstein-Maxwell equations (see Ref.[\refcite{BBB}] and the appendix in Ref.[\refcite{BPP}]). This method has the advantage of a clearer physical interpretation, and it is also straightforward in the calculations; however in the following we will give only the main passages.

 Let there be a central body of mass $m_{in}$ and charge $e_{in}$ and let a spherical massive charged shell with charge $e$ move outside this body. It is clear in advance that the field internal to the shell will be RN, while externally we will have again a RN metric but with different mass and charge parameters $m_{out}$ and $e_{out}=\ei+e$. Using the coordinates $x^0=ct$ and $r$, which are continuous when passing through the shell, we can write the intervals inside, outside, and on the shell as

\begin{align}
	&-(ds)_{in}^{2}=-e^{T(t)} f_{in}(r)c^2dt^2+f_{in}^{-1}(r)dr^2 +r^2d\Omega^2 \label{ds1} \\
  &-(ds)_{out}^{2}=-f_{out}(r)c^2dt^2+f^{-1}_{out}(r) dr^2 +r^2d\Omega^2 \label{ds2}\\
  &-(ds)_{on}^{2}=-c^2d\tau^2+r_0(\tau)^2d\Omega^2  \label{ds3}
\end{align}
where we denoted
\begin{equation}
 d\Omega^2=d\theta^2+\sin^2\theta d\phi^2     \nonumber
\end{equation}
and
\begin{equation}\label{4}
      \fin=1-2\frac{G\,m_{in}}{c^2r}+\frac{G e_{in}^2}{c^4r^2}\,,\ \ \          
      \fo=1-2\frac{G\,m_{out}}{c^2r}+\frac{G(\ei+e)^2}{c^4r^2}.
\end{equation}
In the interval (\ref{ds3}), $\tau$ is the proper time of the shell. The ``dilaton'' factor $e^{T(t)}$ in (\ref{ds1}) is required to ensure the continuity of the time coordinate $t$ through the shell. If the equation of motion for the shell is 
\begin{equation}\label{rR}
      r=R_0(t), 
\end{equation}
then joining the angular part of the three intervals (\ref{ds1})-(\ref{ds3}), one has
\begin{equation}\label{rR2}
      r_0(\tau)=R_0[t(\tau)],
\end{equation}
where the function $t(\tau)$ describes the relationship between the global time and the proper time of the shell. Joining the radial-time parts of the intervals (\ref{ds1})-(\ref{ds2}) on the shell requires that the following relations hold:
\begin{equation}	
\fin(r_0)\left(\dt\right)^2e^{T(t)}-\fin^{-1}(r_0)\left(\dr\right)^2=1\,, \label{6}
\end{equation}
\begin{equation}
\fo(r_0)\left(\dt\right)^2-\fo^{-1}(r_0)\left(\dr\right)^2=1\,.  \label{7}
\end{equation}
If the equation of motion for the shell ---i.e. the function $r_0(\tau)$--- is known, then the function $t(\tau)$ follows from (\ref{7}) and consequently $T(t)$ can be deduced by (\ref{6}). Thus the problem consist only in determining $r_0(\tau)$, which can be done by direct integration of the Einstein-Maxwell equations 
\begin{equation}
\left\{\begin{array}{l}\label{A.1}
        R_i^k-\frac{1}{2}R g_i^k=\frac{8\pi G}{c^4}T_i^k\\\\
        (\sqrt{-g}F^{ik})_{,k}=\sqrt{-g}\frac{4\pi}{c} \rho u^{i}
\end{array}
\right.
\end{equation}
with the energy-momentum tensor given by:
\begin{align}\label{A.2}
	&T_i^k=\epsilon \,u_i u^k+(\delta_i^2\delta_2 ^k+\delta_i^3\delta_3^k)p+T^{(el)}\,_i^k
\\
	&T^{(el)}\,_i^k =\frac{1}{4\pi}(F_{il}F^{kl}-\frac{1}{4} \delta_i^kF_{lm}F^{lm})\ .
\end{align}

 Here on we employ the following notations:
\begin{equation}\nonumber
\begin{array}{ll}
       -ds^2=g_{ik} dx^idx^k , & g_{ik} \ $has signature$\ (-,+,+,+)\\\\
       x^k=(ct,r,\theta,\varphi) &  \ \ i,j,k...=0,1,2,3\\\\
     
       p\equiv p(R_0)= p_{\theta}=p_{\varphi} = $tangential pressure$ & \ \ \ (p_r=0)\\\\
       F_{ik}=A_{k,i}-A_{i,k}
\end{array}
\end{equation}
The above equations are to be solved for the metric
\begin{equation}	\label{8}
-ds^{2}=g_{00}(t,r)c^2dt^2+g_{11}(t,r)dr^2 +r^2d\Omega^2, 
\end{equation}

and for the potential
\begin{equation}	\label{10}
A_0=A_0(t,r),\ \ A_1=A_2=A_3=0,
\end{equation}

As follows from the Landau-Lifshitz approach [\refcite{LL}] (see Ref.[\refcite{BBB}]) the energy distribution of the shell is
\begin{equation}	\label{11}
\ep=\frac{M(t)c^2\delta[r-R_0(t)]}{4\pi r^2 u^0 \sqrt{-g_{00}g_{11}}},
\end{equation}
while its charge density is
\begin{equation}	\label{12}
\rho=\frac{c\,e \delta[r-R_0(t)]}{4\pi r^2u^0\sqrt{-g_{00}g_{11}}},
\end{equation}
where $\delta$ is the standard $\delta$-function. In the absence of tangential pressure $p$, the quantity $M$ in Eqn.(\ref{11}) would be a constant, but in presence of pressure, $Mc^2$ includes the rest energy along with the energy (in the radially comoving frame) of the tangential motions of the particles that produce this pressure.

 It can be checked that the Einstein part of (\ref{A.1}) actually lead to the solution (\ref{ds1})-(\ref{ds3}) with, in addition, the ``joint condition''
\begin{equation}	\label{joint}
\sqrt{\fin(r_0)+\left(\dr\right)^2}+\sqrt{\fo(r_0)+\left(\dr\right)^2}=2\frac{(\mo-\mi)}{\mu(\tau)}
                                                              -\frac{e^2+2ee_{in}}{\mu(\tau)c^2 r_0},
\end{equation}
 where we denoted
\begin{equation}	\label{mu}
\mu(\tau)=M[t(\tau)],
\end{equation} 
while
\begin{equation}	\label{Emo}
\mo-\mi=E/c^2
\end{equation} 
is a constant which can be interpreted as the total amount of energy of the shell.
 Then, from the Maxwell side of (\ref{A.1}) the only non-vanishing component of the electric field is
\begin{equation}	\label{13}
F_{01}=-\frac{\sqrt{-g_{00}g_{11}}}{r^2}\{e_{in}+e\theta[r-R_0(t)]\}
\end{equation}
($\theta(x)$ is the standard step function). Finally, the equations $T^k_{i;k}=0$ can be reduced to the only one relation:
\begin{equation}	\label{14}
p=-\frac{dM}{dt}\frac{c^2\delta[r-R_0(t)]}{8\pi r u^1\sqrt{-g_{00}g_{11}}}
\end{equation}

 We will not treat here the steady case (i.e. $r_0=const$) which should be treated separately; thus in the following we will assume always $r_0\neq const.$ .

 The joint condition (\ref{joint}) can be written in several different forms: two of them, which will be useful in the following, are
 \begin{equation}	\label{18}
\sqrt{\fin(r_0)+\left(\dr\right)^2}=\frac{(\mo-\mi)}{\mu(\tau)}
                                                              +\frac{G\mu^2(\tau)-e^2-2ee_{in}}{2\mu(\tau)c^2 r_0}
\end{equation}
and
 \begin{equation}	\label{19}
\sqrt{\fo(r_0)+\left(\dr\right)^2}=\frac{(\mo-\mi)}{\mu(\tau)}
                                                              -\frac{G\mu^2(\tau)+e^2+2ee_{in}}{2\mu(\tau)c^2 r_0}.
\end{equation}
As in Ref.~\refcite{BBB}, all the radicals encountered here are taken positive, since for astrophysical considerations only these cases are meaningful. To proceed further, we must specify the equation of state, i.e. the function $\mu(\tau)$. Here we consider a particle-made shell, therefore the sum of kinetic and rest energy of all the particles is
\begin{equation}	\label{21}
M c^2=\sum_a \left(m_ac^2\sqrt{1+\frac{p_a^2}{m_a^2c^2}}\right),
\end{equation}
where $p_a$ is the tangential momentum of each particle (the electric interaction between the particles is already taken into account by the self-energy term of, e.g., (\ref{18}), thus one has not to include it in $M$ too). From the definition of the shell (see Introduction) it follows:
\begin{equation}	\label{22}
\frac{p_a^2}{m_a^2}=\frac{l_a^2}{m_a^2R_0^2}=\frac{const}{R_0^2},
\end{equation}
the square root in (\ref{21}) does not depend on the index $a$; then defining
\begin{equation}	\nonumber
\sum_a{m_ac^2}=m c^2,\ \ \ \sum_a{|l_a|}=L,
\end{equation}
formula (\ref{21}) can be re-written (remembering definition (\ref{mu}) too) as
\begin{equation}	\label{24}
\mu(\tau)=\sqrt{m^2+\frac{L^2}{c^2 r_0^2(\tau)}}.
\end{equation}
Thus, now, one can determine the function $r_0(\tau)$ from equation (\ref{joint}) [or from one of the equivalent forms (\ref{18})-(\ref{19})] if the initial radius of the shell and the six arbitrary constants $\mi,\,\mo,\,m,\,e_{in},\,e$ and $L$ are specified.
 Accordingly with (\ref{11}), (\ref{14}), (\ref{mu}) and (\ref{24}), the equation of state  that relates the shell energy density $\ep$ to the tangential pressure $p$ is
\begin{equation}	\label{p}
p=\frac{\ep}{2}\frac{L^2}{m^2c^2R_0^2}\left(1+\frac{L^2}{m^2c^2R^2_0}\right)^{-1}
\end{equation}
as in the uncharged case, i.e. the presence of the charges do not modify the relation between energy density and pressure (indeed the presence of the charge is hidden in the equation of motion). Note that when the shell expands to infinity ($R_0\rightarrow\infty$) the angular momentum becomes irrelevant and the equation of state tends to the dust case $p<<\ep$.

\section{The shells intersection}
\label{inters}
Let us now consider the case of two shells which move in the field of a central charged mass. The generalization from the previous (single-shell) case is straightforward if the shells do not intersect: indeed the outer shell do not affect the motion of the inner one, while the inner one appears from outside just as a RN metric. Therefore the principal aim of this section is to consider the intersection eventuality and to predict the motion of the two shells \textsl{after} the crossing, having specified the initial conditions before the crossing. After the intersection one has a new unknown constant that has to be found by imposing opportune joining conditions as now we are going to explain (the analysis follows step by step the Ref.\refcite{BBB}'s one).

\begin{figure}
\includegraphics[width=9cm, height=7cm]{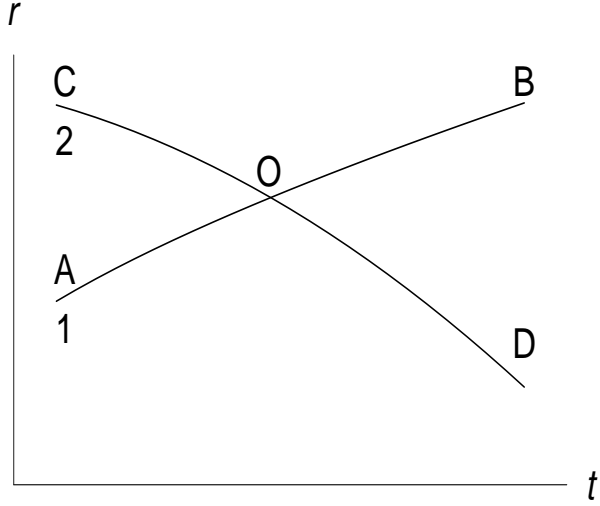}
  \caption{The four region in which it is divided the spacetime; the two lines represent the trajectories of shell-1 and shell-2.}\label{f.1}
\end{figure}

 Let us previously analyze the space-time in the $(t,r)$ coordinates (which are continuous through the shells). We define the point $O\equiv (t_*,r_*)$ as the intersection point; then the space-time is divided in four regions (see Fig.1):
 \begin{equation}	
\begin{array}{lr}
	COB  &  (r>R_1,r>R_2),\\
	COA  &  (R_1<r<R_2),  \\
	AOD  &   (r<R_1,r<R_2), \\
	BOD  & (R_2<r<R_1).	
\end{array}
\end{equation}
Correspondingly to these regions we have the metric in form (\ref{7}) but with different coefficients $g_{00}$ and $g_{11}$: 
\begin{align}
	&g_{00}^{(COB)}=-\fo(r) \,,           &g_{11}^{(COB)}=\fo^{-1}(r) \label{25}\\
	&g_{00}^{(COA)}=-e^{T_1(t)}f_{12}(r)   \,, &g_{11}^{(COA)}=f_{12}^{-1}(r) \label{26}\\
	&g_{00}^{(AOD)}=-e^{T_0(t)}f_{in}(r)\,, &g_{11}^{(AOD)}=f_{in}^{-1}(r) \label{27}\\
	&g_{00}^{(BOD)}=-e^{T_2(t)}f_{21}(r)\,, &g_{11}^{(BOD)}=f_{21}^{-1}(r) \label{28}
\end{align}
The dilaton factor $T(t)$ allows to cover all the space-time with only one $t$-coordinate; here, $\fin$ and $\fo$ are the same as those in (\ref{4}) while $\fot$ and $\fto$ are given by similar expressions:
\begin{align}
	\fot=1-\frac{2G\mot}{c^2r}+\frac{G(\ei+e_1)^2}{c^4r^2} \label{29}\\
	\fto=1-\frac{2G\mto}{c^2r}+\frac{G(\ei+e_2)^2}{c^4r^2} \label{29b}
\end{align}
As we said, the parameters $\mi,\,\mot,\,\mo,\,e_{in},\,e_1,\,e_2$ are assumed to be specified at the beginning, while $\mto$ is the actual unknown constant which has yet to be determined from the joining conditions on $(t_*,r_*)$.

\subsection*{Before the intersection}
\label{sec:BeforeTheIntersection}
Let us write the equation of motion for the two shells before the intersection (shell-1 inner and shell-2 outer). This can be made easily adapting the (\ref{19}) and (\ref{18}) to the present case:
\begin{equation}	\label{30}
\sqrt{\fot(r_1)+\left(\dro\right)^2}=\frac{(\mot-\mi)}{M_1}
                                                              -\frac{GM_1^2+e_1^2+2e_{in}e_1}{2M_1c^2 r_1}
\end{equation}
for shell 1, while for shell 2
 \begin{equation}	\label{31}
\sqrt{\fot(r_2)+\left(\drt\right)^2}=\frac{(\mot-\mi)}{M_2}
                                                              +\frac{GM_2^2-e_2^2-2(e_{in}+e_1)e_2}{2M_2c^2 r_2}
\end{equation}
with
\begin{equation}	\label{32}
M_1=\sqrt{m_1^2+\frac{L_1^2}{c^2 r_1^2}}, \ \ \ M_2=\sqrt{m_2^2+\frac{L_2^2}{c^2 r_2^2}}.
\end{equation}
Here, $\tau_1$ and $\tau_2$ are the proper times of the first and second shells respectively, while $r_1(\tau_1)=R_1[t(\tau_1)]$ and $r_2(\tau_2)=R_2[t(\tau_2)]$. Now we have to impose the joining conditions for the intervals on both the shells. For the first shell (on curve AO) one has:
\begin{equation}	\label{33}
e^{T_1(t)}\fot(r_1)\left(\dto\right)^2-\fot^{-1}(r_1)\left(\dro\right)^2=1
\end{equation}
\begin{equation}	\label{34}
e^{T_0(t)}\fin(r_1)\left(\dto\right)^2-\fin^{-1}(r_1)\left(\dro\right)^2=1;
\end{equation}
while for the second shell:
\begin{equation}	\label{35}
\fo(r_2)\left(\dtt\right)^2-\fo^{-1}(r_2)\left(\drt\right)^2=1
\end{equation}
\begin{equation}	\label{36}
e^{T_1(t)}\fot(r_2)\left(\dtt\right)^2-\fot^{-1}(r_2)\left(\drt\right)^2=1.
\end{equation}

If all free parameters and initial data to Eqs.(\ref{30})-(\ref{32}) were specified and if the functions $r_1(\tau_1)$ and $r_2(\tau_2)$ were derived, then their substitution in (\ref{33})-(\ref{36}) gives the functions $\tau_1(t)$, $\tau_2(t)$ and $T_1(t)$, $T_0(t)$, which is enough for determining the motion of the shells before the intersection. Therefore the intersection point $(t_*,r_*)$ can be found by solving the system
\begin{equation}	\label{37}
\left\{\begin{array}{l}
	r_*=r_1(\tau_1(t_*))\\
	r_*=r_2(\tau_2(t_*))\ ,
\end{array}
\right.
\end{equation}
which we assume that has a solution.

\subsection*{After the intersection}
\label{sec:AfterTheIntersection}
The equation of motion for the shells after the intersection time $t_*$ can be constructed in the same way again by turning to Eqns.(\ref{18}) and (\ref{19}), and introducing the new parameter $\mto$ which characterize the ``Schwarschild mass'' seen by an observer in the region $BOD$. We use Eq.(\ref{18}) for  (now outer) shell 1 and Eq.(\ref{19}) for (now inner) shell 2:
\begin{equation}	\label{38}
\sqrt{\fto(r_1)+\left(\dro\right)^2}=\frac{(\mo-\mto)}{M_1}
                                                              +\frac{GM_1^2-e_1^2-2e_1(e_{in}+e_2)}{2M_1c^2 r_1} \ ,
\end{equation}

\begin{equation}	\label{39}
\sqrt{\fto(r_2)+\left(\drt\right)^2}=\frac{(\mto-\mi)}{M_2}
                                                             -\frac{GM_2^2+e_2^2+2e_2e_{in}}{2M_2c^2 r_2} \ .
\end{equation}
Naturally, $M_1(r_1)$ and $M_2(r_2)$ are given by the same expression of (\ref{32}) but now they have to be calculated on $r_1(\tau_1)$ and $r_2(\tau_2)$ after the intersection.

 Joining the intervals on the first shell (on curve $OB$) yields
\begin{equation}	\label{40}
\fo(r_1)\left(\dto\right)^2-\fo^{-1}(r_1)\left(\dro\right)^2=1
\end{equation}
\begin{equation}	\label{41}
e^{T_2(t)}\fto(r_1)\left(\dto\right)^2-\fto^{-1}(r_1)\left(\dro\right)^2=1.
\end{equation}
Then, joining the second shell (on curve $OB$) we obtain:
\begin{equation}	\label{42}
e^{T_2(t)}\fto(r_2)\left(\dtt\right)^2-\fto^{-1}(r_2)\left(\drt\right)^2=1
\end{equation}
\begin{equation}	\label{43}
e^{T_0(t)}\fin(r_2)\left(\dtt\right)^2-\fin^{-1}(r_2)\left(\drt\right)^2=1.
\end{equation}
Since the initial data to Eqs.(\ref{38}) and (\ref{39}) have already been specified (from the previous evolution), then the evolution of the shells after the intersection would be determined from Eqs.(\ref{38})-(\ref{43}) if parameter $\mto$ were known.
Thus we need an additional physical condition from which we could determine $\mto$.

 This condition follows from the fact that the Christoffel symbols (i.e. the accelerations) of the shells have only finite discontinuities (finite jumps), therefore the relative velocity of the shells must remain continuous through the crossing point.

 In the presence of two shells, we can construct one more invariant than in the single shell case (where only $u_i u^i=-1$ was possible): the scalar product between the two 4-velocities of the shells. We can also avoid to apply the parallel transport if we evaluate the 4-velocities on the intersection point $(t_*,r_*)$. The continuity condition can be found imposing that the scalar product has to have the same value when evaluated in both the two limits $t\rightarrow t_*^-$ and $t\rightarrow t_*^+$.

\subsection*{Determination of Q.}
Let us start determining the quantity
\begin{equation}	\label{52}
Q\equiv \{g_{00}^{(COA)}u^0_{AO}u^0_{CO}+g_{11}^{(COA)}u^1_{AO}u^1_{CO}\}_{t=t^*,r=r_1=r_2=r_*},
\end{equation}
which is the scalar product of the two 4-velocities evaluated in the intersection point from the region $AOC$ (along the curves $AO$ and $CO$).
Written explicitly, the unit tangent vector to trajectory $AO$ is

\begin{align}	
	u^i_{AO}&=(u^0_{AO},u^1_{AO},u^2_{AO},u^3_{AO}) \nonumber \\
          &=\left(\dto,\dro,0,0\right)_{t\leq t_*}, \label{44}
\end{align}

  while for the trajectory $CO$ we have
	
\begin{align}
	u^i_{CO}&=(u^0_{CO},u^1_{CO},u^2_{CO},u^3_{CO})\nonumber\\
          &=\left(\dtt,\drt,0,0\right)_{t\leq t_*}.\label{45}
\end{align}

The fact that these are actually unit vectors follows from the joining equations (\ref{33}) and (\ref{36}).

 The components of the vector (\ref{44}) can be easily expressed from Eqs.(\ref{30}) and (\ref{33}) as
\begin{equation}	\label{46}
\left(\dto\right)_{t\leq t_*}=\frac{e^{-T_1(t)/2}}{M_1(r_1)\fot(r_1)}\left(\mot-\mi-\frac{GM_1^2(r_1)+e_1^2+2e_1e_{in}}{2c^2r_1}\right)
\end{equation}

\begin{align}
\left(\dro\right)_{t\leq t_*}&= \nonumber\\
 =&\frac{\delta_1}{M_1(r_1)\fot(r_1)}\sqrt{\left(\mot-\mi-\frac{GM_1^2(r_1)+e_1^2+2e_1e_{in}}{2c^2r_1}\right)^2-M_1^2(r_1)\fot(r_1)}	\label{47}
 \end{align}

where
\begin{equation}	\label{48}
\delta_1=\texttt{sgn}\left(\dro\right)_{t\leq t_*}.
\end{equation}
Analogously, for the components of vector (\ref{45}), we obtain the following expressions from Eqs.(\ref{31}) and (\ref{36}): 
\begin{equation}	\label{49}
\left(\dtt\right)_{t\leq t_*}=\frac{e^{-T_1(t)/2}}{M_2(r_2)\fot(r_2)}\left(\mo-\mot+\frac{GM_2^2(r_2)-e_2^2-2e_2(e_{in}+e_1)}{2c^2r_2}\right)
\end{equation}

\begin{align}
\left(\drt\right)_{t\leq t_*}&= \frac{\delta_2}{M_2(r_2)\fot(r_2)}\, \cdot \nonumber \\
 \cdot \, & \sqrt{\left(\mo-\mot+\frac{GM_2^2(r_2)-e_2^2-2e_2(e_{in}+e_1)}{2c^2r_2}\right)^2-M_2^2(r_2)\fot(r_2)}	\label{50}
 \end{align}

\begin{equation}	\label{51}
\delta_2=\texttt{sgn}\left(\drt\right)_{t\leq t_*}.
\end{equation}
Thus, from the preceding results, we obtain:
 \begin{equation}	\label{53}
\begin{array}{l l}
Q=&\frac{-1}{M_1M_2\fot}\,\cdot\\\\
       &\,\cdot\Bigl\{

                      \left(\mot-\mi-\frac{GM_1^2+e_1^2+2e_1e_{in}}{2c^2r_*}\right)
                      \left(\mo-\mot+\frac{GM_2^2-e_2^2-2e_2(e_{in}+e_1)}{2c^2r_*}\right)+\\\\
  
  & -\delta_1\delta_2\sqrt{\left(\mot-\mi-\frac{GM_1^2+e_1^2+2e_1e_{in}}{2c^2r_*}\right)^2-M_1^2\fot}
\\\\
  & \sqrt{\left(\mo-\mot+\frac{GM_2^2-e_2^2-2e_2(e_{in}+e_1)}{2c^2r_*}\right)^2-M_2^2\fot} \Bigl\};

 \end{array}
\end{equation}
here and in the following we omit the coordinate dependence of $f_a$, $M_a$ etc., implicitly assuming that they have to be evaluated on $(t_*,r_*)$ where not differently indicated.

\subsection*{Determination of Q$'$.}
\label{sec:DeterminationOfQ}
It is possible to apply the same procedure to the region $BOD$ (i.e. after the intersection time), finding the quantity
\begin{equation}	\label{56}
Q'\equiv \{g_{00}^{(BOD)}u^0_{OB}u^0_{OD}+g_{11}^{(BOD)}u^1_{OB}u^1_{OD}\}_{t=t^*,r=r_1=r_2=r_*}\ .
\end{equation}

Now the unit tangent vectors to trajectories $OB$ and $OD$ are\footnote{Obviously, when we say $t\geq t_*$, we tacitly assume before a (possible) second intersection.}:

\begin{align}
	u^i_{OB}&=(u^0_{OB},u^1_{OB},u^2_{OB},u^3_{OB}) \nonumber\\
          &=\left(\dto,\dro,0,0\right)_{t\geq t_*},	\label{54}
\end{align}

  and

\begin{align}
	u^i_{OD}&=(u^0_{OD},u^1_{OD},u^2_{OD},u^3_{OD})\nonumber\\
          &=\left(\dtt,\drt,0,0\right)_{t\geq t_*};	\label{55}
\end{align}

from the joining conditions (\ref{41}) and (\ref{42}) it is possible to see that these are actually unit vectors. The components of $u^i_{OB}$ can be deduced from Eqs.(\ref{38}) and (\ref{41}), while the components of $u^i_{OD}$ from Eqs.(\ref{39}) and (\ref{42}). Then, using the metric in the region $BOD$, it is possible to calculate the scalar product
\begin{equation}	\label{56b}
\begin{array}{l l}
Q'=&\frac{-1}{M_1M_2\fto}\,\cdot\\\\
       &\,\cdot\Bigl\{

                      \left(\mo-\mto+\frac{GM_1^2-e_1^2-2e_1(e_{in}+e_2)}{2c^2r_*}\right)
                      
                      \left(\mto-\mi-\frac{GM_2^2+e_2^2+2e_2e_{in}}{2c^2r_*}\right)+\\\\
  
  & -\delta_1'\delta_2'\sqrt{\left(\mo-\mto+\frac{GM_1^2-e_1^2-2e_1(e_{in}+e_2)}{2c^2r_*}\right)^2-M_1^2\fto}
\\\\
  & \sqrt{\left(\mto-\mi-\frac{GM_2^2+e_2^2+2e_2e_{in}}{2c^2r_*}\right)^2-M_2^2\fto}\, \Bigl\} \ ,

 \end{array}
\end{equation}
where $\delta_1'$ and $\delta_2'$ have been defined as in (\ref{48}) and (\ref{51}), but for $t\geq t_*$. We introduced these symbols only for generality, but actually we are interested only in the case with\footnote{This is the only possible case if one excludes $v_1(t^*)=v_2(t^*)=0$, because there are non discontinuities in the velocities.}  

\begin{equation}
	\delta_1'= \delta_1,\ \ \ \ \delta_2'= \delta_2 \ .
\end{equation}

 The necessary continuity requirement is thus
 \begin{equation}	\label{58}
 Q=Q',
\end{equation}
 then, since $r_*$ is assumed to be known, this equation allows to find $\mto$.


\subsection*{Physical meaning of Q and Q$'$.}
\label{sec:PhysicalInterpretationOfQAndQ}
Using standard definition for the shell velocities before the intersection one has
\begin{equation}	\label{59}
\left(\frac{v_1}{c}\right)^2=\frac{g_{11}^{(COA)}(r_1)}{-g_{00}^{(COA)}(r_1)}\left(\drot\right)^2
\end{equation}

\begin{equation}	\label{59.b}
\left(\frac{v_2}{c}\right)^2=\frac{g_{11}^{(COA)}(r_2)}{-g_{00}^{(COA)}(r_2)}\left(\drtt\right)^2\ ,
\end{equation}
and similarly for the velocities after the intersection,
\begin{equation}	\label{63}
\left(\frac{v_1'}{c}\right)^2=\frac{g_{11}^{(BOD)}(r_1)}{-g_{00}^{(BOD)}(r_1)}\left(\drot\right)^2
\end{equation}

\begin{equation}	\label{63.b}
\left(\frac{v_2'}{c}\right)^2=\frac{g_{11}^{(BOD)}(r_2)}{-g_{00}^{(BOD)}(r_2)}\left(\drtt\right)^2\ .
\end{equation}
Then it is easy to obtain from the definitions (\ref{52}) and (\ref{56}), that\footnote{It is also worth noting that $\sqrt{Q^2-1}/Q=-|v_1/c-v_2/c|/(1-v_1v_2/c^2)$, which is the relative velocity definition of two ``particles'' in relativistic mechanics.} 
\begin{equation}	\label{62}
Q=\left\{\frac{v_1 v_2/c^2-1}{\sqrt{1-v_1^2/c^2}\sqrt{1-v_2^2/c^2}}\right\}_{t=t_*,r_1=r_2=r_*}
\end{equation}
and
\begin{equation}	\label{66}
Q'=\left\{\frac{v_1' v_2'/c^2-1}{\sqrt{1-(v_1')^2/c^2}\sqrt{1-(v_2')^2/c^2}}\right\}_{t=t_*,r_1=r_2=r_*}.
\end{equation}

\subsection*{Determination of P and P$'$.}
\label{sec:DeterminationOfPAndP}
First of all it is convenient to introduce new symbols to simplify the expressions of Q and Q'. With

\begin{align}	
	q_1&\equiv -\frac{GM_1^2+e_1^2+2e_1e_{in}}{2c^2r_*}  \nonumber\\
	q_2&\equiv \frac{GM_2^2-e_2^2-2e_2(e_{in}+e_1)}{2c^2r_*}\nonumber \,,
\end{align}

and

\begin{align}
	q_1'&\equiv \frac{GM_1^2-e_1^2-2e_1(e_{in}+e_2)}{2c^2r_*} \nonumber \\
	q_2'&\equiv -\frac{GM_2^2+e_2^2+2e_2e_{in}}{2c^2r_*}\nonumber\,,
\end{align}

then $Q$ and $Q'$ can be re-written as
 \begin{equation}	\label{70}
\begin{array}{l l}
Q=&\frac{-1}{M_1M_2\fot}\,\cdot\\\\
       &\,\cdot\Bigl\{

                      \left(\mot-\mi+q_1\right)
                      \left(\mo-\mot+q_2\right)+\\\\
  
  & -\delta_1\delta_2\sqrt{\left(\mot-\mi+q_1\right)^2-M_1^2\fot}
\\\\
  & \sqrt{\left(\mo-\mot+q_2\right)^2-M_2^2\fot}\, \Bigl\} \,

 \end{array}
\end{equation}
and
\begin{equation}	\label{71}
\begin{array}{l l}
Q'=&\frac{-1}{M_1M_2\fto}\,\cdot\\\\
       &\,\cdot\Bigl\{

                      \left(\mo-\mto+q_1'\right)
                      
                      \left(\mto-\mi+q_2'\right)+\\\\
  
  & -\delta_1'\delta_2'\sqrt{\left(\mo-\mto+q_1'\right)^2-M_1^2\fot}
\\\\
  & \sqrt{\left(\mto-\mi+q_2'\right)^2-M_2^2\fot}\, \Bigl\} \ ,

 \end{array}
\end{equation}
Now, in principle is possible to find $\mto$ by squaring  and solving $Q=Q'$ (which is a quartic equation). However the procedure is cumbersome and moreover it is not possible with Eq.(\ref{58}) alone to determine the sign of the roots. Fortunately, as in the non-charged case, it is possible to follow another easier way. Indeed, it is possible to introduce two other invariants, say $P$ and $P'$, similar to $Q$ and $Q'$, which are constructed using the scalar products of the 4-velocities of the two shell, but now taking the limit to $(t_*,r_*)$ from the $AOD$ and $COB$ regions respectively. More explicitly, we define
\begin{equation}	\label{P}
P\equiv \{g_{00}^{(AOD)}u^0_{AO}u^0_{OD}+g_{11}^{(AOD)}u^1_{AO}u^1_{OD}\}_{t=t^*,r=r_1=r_2=r_*} \ ,
\end{equation}
and
\begin{equation}	\label{Pp}
P'\equiv \{g_{00}^{(COB)}u^0_{CO}u^0_{OB}+g_{11}^{(COB)}u^1_{CO}u^1_{OB}\}_{t=t^*,r=r_1=r_2=r_*} \ .
\end{equation}
Then, the same continuity requirement of Eq.(\ref{58}) implies that it must hold also that
\begin{equation}	\label{74}
Q=P\  ,\ \ \ \ \ P=P'\ .
\end{equation}
Following the same method used to find $Q$ and $Q'$, after some calculations, one arrives to
\begin{equation}	\label{75}
\begin{array}{l l}
P=&\frac{-1}{M_1M_2\fin}\,\cdot\\\\
       &\,\cdot\Bigl\{

                      \left(\mot-\mi+p_1\right)
                      \left(\mto-\mi+p_2\right)+\\\\
  
  & -\delta_1\delta_2'\sqrt{\left(\mot-\mi+p_1\right)^2-M_1^2\fin}
\\\\
  & \sqrt{\left(\mto-\mi+p_2\right)^2-M_2^2\fin}\, \Bigl\}

 \end{array}
\end{equation}
and
\begin{equation}	\label{76}
\begin{array}{l l}
P'=&\frac{-1}{M_1M_2\fin}\,\cdot\\\\
       &\,\cdot\Bigl\{

                      \left(\mo-\mto+p_1'\right)
                      \left(\mo-\mot+p_2'\right)+\\\\
  
  & -\delta_1'\delta_2\sqrt{\left(\mo-\mto+p_1'\right)^2-M_1^2\fo}
\\\\
  & \sqrt{\left(\mo-\mot+p_2'\right)^2-M_2^2\fo}\, \Bigl\}  \ ,

 \end{array}
\end{equation}
where we have denoted

\begin{align}
	p_1&\equiv \frac{GM_1^2-e_1^2-2e_1e_{in}}{2c^2r_*}  \nonumber\\
	p_2&\equiv \frac{GM_2^2-e_2^2-2e_2e_{in}}{2c^2r_*}\ ,\nonumber
\end{align}

and

\begin{align}
	p_1'&\equiv -\frac{GM_1^2+e_1^2+2e_1(e_{in}+e_2)}{2c^2r_*}  \nonumber\\
	p_2'&\equiv -\frac{GM_2^2+e_2^2+2e_2(e_{in}+e_1)}{2c^2r_*}\ .\nonumber
\end{align}

\subsection*{Determination of $\mto$; the energy transfer}
\label{sec:DeterminationOfMto}
Thus the complete set of continuity conditions at the point of intersection can be written as
\begin{equation}	\label{77}
Q=Q',\ \ \ Q=P,\ \ \ Q=P'.
\end{equation}
It turns out that this three quartic equations for the unknown parameter $\mto$ have only one common root\footnote{In the pure gravitational case it is also possible to use just Eqn. $Q=P'$, and then choose the correct root solution by requiring that $m_{12}$ has to be positive, see Ref.~\refcite{LE}.}
. It is possible to find the solution using hyperbolic functions.
The final result is remarkably simple:
\begin{equation}	\label{78}
\mto=\mi+\mo-\mot-\frac{e_1e_2}{c^2r_*}-\frac{GM_1M_2}{c^2r_*}Q \ ,
\end{equation}
or equivalently, in terms of $\fto$:
\begin{equation}	\label{79}
\fto=\fin+\fo-\fot+2\frac{G^2M_1M_2}{c^4r_*^2}Q \ .
\end{equation}
It can be easily seen from Eqn.(\ref{78})  that the charge $e_{in}$ of the central singularity does not affect the result (but it affects the equation of the motion of the shells and thus $Q$).
Formula (\ref{78}) solves the problem of determining the mass parameter $\mto$ from the quantities specified at the evolutionary stage before intersection.
 It is then possible to determine the energy transfer between the shells. Indeed the energy of shell 1 and 2 before the intersection are, respectively
 \begin{equation}	\label{79E}
E_1=(\mot-\mi)c^2\ ,\ \ \ E_2=(\mo-\mot)c^2\ ,
\end{equation}
while, after the intersection
 \begin{equation}	\label{80}
E_1'=(\mo-\mto)c^2\ ,\ \ \ E_2'=(\mto-\mi)c^2 \ .
\end{equation}
The conservation of total energy is automatically ensured by the above formulas, indeed
\begin{equation}	\label{81}
E_1+E_2=E_1'+E_2'\ .
\end{equation}
Then it is natural to define the exchange energy as
\begin{equation}	\label{82}
\Delta E=E_2'-E_2=-(E_1'-E_1) \ .
\end{equation}
Then, from Eqn.(\ref{78})  and the above definitions, it follows that
\begin{equation}	\label{83}
\Delta E=-\frac{e_1e_2}{r_*}-\frac{GM_1M_2}{r_*}Q \ .
\end{equation}
It is also useful (especially for the Newtonian approximation) to use Eqn.(\ref{62}) and re-express $\Delta E$ as:
\begin{equation}	\label{84}
\Delta E =-\frac{e_1e_2}{r_*}
        -\frac{GM_1M_2}{r_*}\left\{\frac{v_1 v_2/c^2-1}{\sqrt{1-v_1^2/c^2}\sqrt{1-v_2^2/c^2}}\right\}_{r=r_*} \ .
\end{equation}
\section{Post-Newtonian approximation}
\label{sec:NewtonianApproximation}
For slow velocities of the shells it is interesting to consider the Post-Newtonian limit of Eqn.(\ref{84}):
	
\begin{align}
	\Delta E=&\frac{Gm_1m_2-e_1e_2}{r_*}+ \nonumber\\
	         &+\frac{1}{2c^2}\left\{\frac{Gm_1m_2}{r_*}[v_1(r_*)-v_2(r_*)]^2+\frac{Gm_2L_1^2}{m_1r_*^3}+ \frac{Gm_1L_2^2}{m_2r_*^3} \right\}+o\left(\frac{1}{c^4}\right) \ .\label{99}
\end{align}

It is worth noting that only the zeroth order in $1/c^2$ changes with respect to the uncharged case (because of the Coulomb term $-e_1e_2/r_*$), while all the other orders remain unchanged, being of kinetic origin; $m_1$ and $m_2$ are the rest masses of the shells, indeed  we have used for the masses $M_1$ and $M_2$ the definitions (\ref{32}). 

 It can be also useful to re-express all the quantities in a Newtonian language and consider only the zeroth order in $1/c^2$, e.g. we can expand the energy as
\begin{equation}	\label{E}
E=mc^2+\mathcal E+o\left(\frac{1}{c^2}\right)\ ,
\end{equation}
where $m$ and $\mathcal E$ do not depend on $c$. Therefore, similarly, we can define at the first order in $1/c^2$
\begin{align}	\label{87}
\mot-\mi=m_1+\frac{\mathcal E_1}{c^2}, \ \ \ \ \mo-\mot=m_2+\frac{\mathcal E_2}{c^2},\\
\mo-\mto=m_1+\frac{\mathcal E_1'}{c^2}, \ \ \ \ \mto-\mi=m_1+\frac{\mathcal E_2'}{c^2}.
\end{align}
Then it follows also that the energy conservation law takes the form
\begin{equation}	\label{89}
\mathcal E_1+\mathcal E_2=\mathcal E_1'+\mathcal E_2' \  ,
\end{equation}
and Eqn.(\ref{82}) becomes
\begin{equation}	\label{90}
\mathcal E_1'=\mathcal E_1-\Delta \mathcal E,\ \ \ \ \mathcal E_2'=\mathcal E_2+\Delta \mathcal E\,,
\end{equation}
where $\Delta\mathcal E=(\Delta E)_{c\rightarrow\infty}$. Thus from the above formulas and definitions it is clear that
\begin{equation}	\label{91}
\Delta \mathcal E=\frac{\ga}{r_*} \ .
\end{equation}


\section{Pressureless shells with zero effective masses ($L_1=L_2=0$ and $M_1=M_2=0$)}
\label{sec:ShellWithZeroEffectiveMasses}
It is interesting also to consider the case in which the motion of the particles of the shells is only radial (i.e. $L_1=L_2=0$) and the rest masses are negligible with respect to the kinetic energies and to the charges ---indeed this is the case for two shells composed by (ultra)rela-tivistic electrons and positrons. In this case the effective masses can be replaced by 
\begin{equation}
	M_1=M_2=\lambda\,,
\end{equation}
where $\lambda$ is a parameter arbitrary small.
 From Eqn.(\ref{83}), with $Q$ expressed by formula (\ref{53}), it is easy to find that the energy transfer in this case is
 \begin{align}\label{120}
	\Delta E=-\frac{e_1e_2}{r_*}+\frac{c^4 r_*}{2G\fot}(\fin-\fot)(\fot-\fo)+o(\lambda^2)\,,
\end{align}
 having assumed that the shells have opposite-directed velocities, i.e.
\begin{align}\label{dd}
	\delta_1\delta_2=-1\ .
\end{align}
Otherwise, if the shells goes in the same direction, i.e.
\begin{align}\label{dd1}
	\delta_1\delta_2=1\ ,
\end{align}
then Eqn.(\ref{83}) becomes simply
 \begin{align}\label{120l}
	\Delta E=-\frac{e_1e_2}{r_*}+o(\lambda^2)\ ;
\end{align}
obviously the previous formulas make sense only if $r_*$ exists.
We want to underline the presence of the term $o(\lambda^2)$, because, strictly speaking, a charge cannot have zero rest mass, therefore we are in the case of just \textsl{small} effective masses. As expected, in the case of vanishing charges ($e_1=e_2=0$), Eqn.(\ref{120l}) gives zero at $\lambda=0$ because this is the case of two photon-shells which go in the same direction and therefore cannot never intersect.

\section{The intersection of a test shell with a gravitating one}
\label{sec:TheIntersectionOfATestShellWithAGravitatingOne}

\subsection*{One-shell case}
\label{sec:OneShellCase}
Let us consider firstly the case of a test shell on the RN field. This limit has the only aim to show that the shell's equation of motion (\ref{18}) actually reduce to the simple test-particle case; the limit can be obtained by putting
 \begin{align}\label{l}
	m \rightarrow\lambda m\ ,\ \  	e \rightarrow\lambda m \ , \ \ 	L \rightarrow\lambda L \ ,\ \ 	(\mo-\mi)c^2 \rightarrow\lambda E
\end{align}
with $\lambda\rightarrow 0$. Then, considering also (\ref{24}), we find that Eqn.(\ref{18}) becomes
 \begin{align}\label{128}
E=\mu c^2\sqrt{\fin(r_0)+\left(\dr\right)^2}+\frac{e\ei}{r_0}-\lambda \frac{G\mu^2-e^2}{2r_0} \ ,
\end{align}
now, putting $\lambda =0$ the self-energy term is killed; then re-writing Eqn.(\ref{128}) using the more familiar Schwarzschild time $t$ and Eqn.(\ref{rR2}),
\begin{align}\label{129}
E=c^2\sqrt{m^2+\frac{L^2}{c^2 R_0^2(t)}} \sqrt{\frac{\fin^3(R_0)}{\fin^2(R_0)-\left(\frac{dR_0}{cdt}\right)^2}}+\frac{e\ei}{R_0}+o(\lambda)\  ,
\end{align}
it is easy to recognize that Eqn.(\ref{129}) coincides with the first integral of motion of a test-charge particle on the Reinssner-Nordstrom background, where $E$ is the conserved energy of the particle, $m$ the rest mass, $e$ the charge and $L$ the angular momentum.
\subsection*{Two-shell case, with one test-shell}
\label{sec:TwoShellCaseWithOneTestShell}
Now we can deal with the more interesting two-shell case, in which shell-2 is considered ``test''. To gain this limit we have to put
 \begin{align}\label{l2}
	m_2 \rightarrow\lambda m_2\ ,\ \  	e_2 \rightarrow\lambda m_2 \ , \ \ 	L_2 \rightarrow\lambda L_2 \ ,\\ \nonumber	(\mo-\mot)c^2 \rightarrow\lambda E_2 \ , \ \ (\mto-\mi)c^2 \rightarrow\lambda E_2' \ .
\end{align}
Then, using Eqn.(\ref{78}) with $Q$ given by formula (\ref{53}), one obtains
 \begin{equation}	\label{137}
\begin{array}{l l}
	\Delta E=&-\frac{e_1e_2}{r_*}+\frac{1}{r_*\fot}\,\cdot\\\\
       &\,\cdot\Bigl\{

                      \left(E_1-\frac{GM_1^2+e_1^2+2e_1e_{in}}{2c^2r_*}\right)
                      \left(E_2-\frac{e_2(e_{in}+e_1)}{c^2r_*}+\lambda\frac{GM_2^2-e_2^2}{2c^2r_*}\right)+\\\\
  
  & -\delta_1\delta_2\sqrt{\left(E_1-\frac{GM_1^2+e_1^2+2e_1e_{in}}{2c^2r_*}\right)^2-M_1^2\fot}
\\\\
  & \sqrt{\left(E_2-\frac{e_2(e_{in}+e_1)}{c^2r_*}+\lambda\frac{GM_2^2-e_2^2}{2c^2r_*}\right)^2-M_2^2\fot}\  \Bigl\}\ .

 \end{array}
\end{equation}
Thus, only the self-energy terms of shell-2 are killed by $\lambda=0$.

Now, it is worth noting the following fact: shell-1 does not have any discontinuity when it intersect the shell-2 (this is natural because shell-2 is ``test'' and does not affect the metric), on the other hand shell-2 undergoes a discontinuity in the metric when it cross shell-1 and consequently it has an actual discontinuity in the velocity. It is easy to calculate this gap; indeed using the definition (\ref{59.b}) of velocity $v_2$ [with the time $\dtt$ given by the joint condition (\ref{36})], with metric coefficient (\ref{26}), and with the help the first integral of motion (\ref{31}), one finds
\begin{equation}\label{138}
v_2^2(r_2)=
  1-\fo(r_2)\left(\frac{E_2}{M_2(r_2)}-\frac{e_2(e_1+\ei)^2}{M_2(r_2)r_2}\right)^{-2}+o(\lambda) \ ,\ \ \  t\leq t_* \ ,
\end{equation}
where we have used $\fot=\fo+o(\lambda)$; in the same way, using (\ref{63.b}), (\ref{42}), (\ref{28}), and (\ref{39}), the velocity $v_2'$ (after the intersection) is
\begin{equation}\label{139}
[v_2'(r_2)]^2=
  1-\fin(r_2)\left(\frac{E_2'}{M_2(r_2)}-\frac{e_2(e_1+\ei)^2}{M_2(r_2)r_2}\right)^{-2}+o(\lambda) \ ,\ \ \ t\geq t_*\ ,
\end{equation}
where $E_2'$ can be expressed in function of $E_2$ with the help of (\ref{137}). From the previous formulas it is clear that in general
\begin{equation}\label{139b}
v_2'(r_*)-v_2(r_*)\neq 0 \ .
\end{equation}
\section{Shell ejection}
\label{sec:ShellEjection}
The exchange in energy of the shells during the intersection makes possible that one initially bounded shell can acquire enough energy to escape to infinity. 

 The shell ejection mechanism can take place also in the Newtonian regime. In this case, from Eqs.(\ref{90})-(\ref{91}) it results that
 \begin{align}\label{140}
	\mathcal E_1'=\mathcal E_1-\frac{\ga}{r_*'}\ ,\ \ \ \ \mathcal E_2'=\mathcal E_2+\frac{\ga}{r_*'}\ ,
\end{align}
and then, after the first intersection
\begin{equation}\label{141}
\begin{cases}
	\mathcal E_1''=\mathcal E_1'+\frac{\ga}{r_*''}=\mathcal E_1-(\ga)\left(\frac{1}{r_*'}-\frac{1}{r_*''}\right)\\\\
	 \mathcal E_2''=\mathcal E_2'-\frac{\ga}{r_*''}=\mathcal E_2+(\ga)\left(\frac{1}{r_*'}-\frac{1}{r_*''}\right),
\end{cases}
\end{equation}
where we have denoted the radius of the first and second intersection with $r_*'$ and $r_*''$ respectively. In the following we will consider only the case 
\begin{align}\label{ga}
	\ga>0 \ ,
\end{align}
this is e.g. the case in which the two shells have opposite charges. Thus, also in the case $\mathcal E_1, \mathcal E_2<0$, if 
\begin{align}\label{r'r''}
r_*''>r_*' \ ,
\end{align}
and if the initial condition were in such a way that $r_*'$ is enough small and  $r_*''$ not too much close to $r_*'$, then it is possible to have $\mathcal E_2''>0$, i.e. the ejection of the second shell.

 Let us now assume that $r_*''>r_*'$, and consider a ``semi-relativistic'' case in which at the first intersection we use the full relativistic formulas\footnote{Remember that $-Q=1+o(1/c^2)$.},
\begin{equation}\label{142}
 \begin{cases}
  E_1'=E_1-\frac{M_1(r_*')M_2(r_*')}{r_*'}(-Q)+\frac{e_1e_2}{r_*'}\\\\ E_2'=E_2+\frac{M_1(r_*')M_2(r_*')}{r_*'}(-Q)-\frac{e_1e_2}{r_*'}\ ,
\end{cases} 
\end{equation}
while at the second intersection we use the Newtonian approximation,

\begin{equation}\label{143}
\begin{cases}
	E_1''&= E_1'+\frac{\ga}{r_*''}                                                 \\
	     &=E_1-\left[\frac{M_1(r_*')M_2(r_*')(-Q)-e_1e_2}{r_*'}-\frac{\ga}{r_*''}\right]  
\\\\		E_2''&= E_2'-\frac{\ga}{r_*''}  \\
       &= E_2+\left[\frac{M_1(r_*')M_2(r_*')(-Q)-e_1e_2}{r_*'}-\frac{\ga}{r_*''}\right]\,.
\end{cases}
\end{equation}
This approximation is always justified if the radius of the second intersection $r_*''$ is enough large. Now, it is remarkable that whatever the value of $r_*'$ is, the first term in the square brackets in Eqn.(\ref{143}) satisfies the inequality
\begin{align}\label{144}
	\frac{M_1(r_*')M_2(r_*')(-Q)-e_1e_2}{r_*'}>\frac{\ga}{r_*'}\ .
\end{align}
Comparing the expressions (\ref{143}), (\ref{144}) and (\ref{141}) it is possible to see that in the relativistic regime the shell ejection possibility is even greater than in the Newtonian case. Furthermore, it is worth noting that the presence of the charge do not change qualitatively the pure gravitational analysis, but just magnifies the ejection effect. 

\subsection{$\ga<0$ case}
\label{sec:Ga0Case}
Let us consider also briefly the case in which the shells are equal-signed charged and the repulsion overcome the gravity attraction, i.e. $\ga<0$. In this case the ejection can happen only after an odd number of intersections. 

 E.g. after three intersections, from the previous formulas we have, in the Newtonian approximation:
\begin{align}\label{145}
\mathcal E_1'''=\mathcal E_1-(\ga)\left(\frac{1}{r_*'}-\frac{1}{r_*''}+\frac{1}{r_*'''}\right) \ .
\end{align}
Obviously this formula has a meaning only if 
\begin{align}\label{146}
\frac{1}{r_*'}<\frac{1}{r_*'}-\frac{1}{r_*''}+\frac{1}{r_*'''} \ ,
\end{align}
otherwise the ejection happens at the first intersection (and then there would not be other crossings, and no $r_*'',\ r_*'''$), or never more; if Eqn.(\ref{146}) is true, then it means that the barycenter of the two shells is falling into the center singularity.
\section{Conclusions}
\label{sec:Conclusions}
We have found the energy exchange between two charged crossing shells (formula (\ref{84})). Then we have studied special cases of physical interest in which the formulas simplify: the non relativistic case, the massless shells, the test shell, and finally the ejection mechanism in a semi-Newtonian regime: we found that the ejection mechanism is more efficient in the charged case than in the neutral one if the charges have opposite sign (because the energy transfer is larger due to the Coulomb interaction).

\bibliographystyle{unsrt}

\end{document}